\documentclass[aps,showpacs,twocolumn,floatfix,bibnotes]{revtex4}

\usepackage{here}

\usepackage[dvips]{graphicx}

\newcommand\pictc[5]{\begin{figure}
                       \centerline{
                       \includegraphics[width=#1\columnwidth]{#3}}
                   \protect\caption{\protect\label{fig:#4} #5}
                    \end{figure}            }
\newcommand\pict[4][1.]{\pictc{#1}{!tb}{#2}{#3}{#4}}
\newcommand\rpict[1]{\ref{fig:#1}}

\newcommand\leqt[1]{\protect\label{eq:#1}}
\newcommand\reqtn[1]{\ref{eq:#1}}
\newcommand\reqt[1]{(\reqtn{#1})}

\newcounter{Fig}

\begin{document}
\begin{sloppy}

\title{Transmission properties of left-handed band-gap structures}

\author{Ilya V. Shadrivov$^1$}
\author{Nina A. Zharova$^{1,2}$}
\author{Alexander A. Zharov$^{1,3}$}
\author{Yuri S. Kivshar$^1$}

\affiliation{$^1$ Nonlinear Physics Group and Centre for Ultra-high
bandwidth Devices for Optical Systems (CUDOS), \\
Research School of Physical Sciences and Engineering, \\
Australian National University, Canberra ACT 0200, Australia \\
$^2$ Institute of Applied Physics, Russian Academy of Sciences, Nizhny
 Novgorod 603600, Russia \\
$^3$ Institute for Physics of Microstructures, Russian Academy of
Sciences, Nizhny Novgorod 603950, Russia }

\begin{abstract}
We analyze transmission of electromagnetic waves through a periodic band-gap structure consisting of slabs of a left-handed metamaterial and air. Using the effective parameters of the metamaterial derived from its microscopic
structure, we study, with the help of the transfer-matrix approach
and by means of the finite-difference-time-domain numerical simulations, the
transmission properties of such a left-handed photonic crystals in a novel type of band gap associated with the zero averaged refractive index. We demonstrate that the transmission can be made tunable by introducing defects, which allow to access selectively two different types of band gaps.
\end{abstract}


\maketitle

\section{Introduction}

Materials with both negative dielectric permittivity and negative magnetic
permeability were suggested theoretically a long time
ago~\cite{Veselago:1967-517:UFN}, and they are termed {\em left-handed
materials} because the wave vector and Poynting vector lie in the
opposite directions. Many unusual properties of such materials can
be associated with their {\em negative refractive index}, as was demonstrated by several reliable experiments~\cite{Shelby:2001-77:SCI,Parazzoli:2003-107401:PRL} and numerical finite-difference-time-domain (FDTD) simulations (see, e.g., Ref.~\cite{Foteinopoulou:2003-107402:PRL}).

Multilayered structures that include materials with negative refraction can be considered as a sequence of the flat lenses that provide an optical cancellation of the layers with positive refractive index leading to either enhanced or suppressed transmission~\cite{Zhang:2002-1097:APL,Nefedov:2002-36611:PRE, Pendry:2003-6345:JPCM}. More importantly, a one-dimensional stack of layers with alternating positive and
negative-refraction materials with {\em zero averaged refractive index} displays a novel type of the transmission
band gap~\cite{Nefedov:2002-36611:PRE, Li:2003-83901:PRL, Ruppin:2003-494:MOTL, Shadrivov:2003-3820:APL, Wu:2003-235103:PRB} near the frequency
where the condition $\mathopen{<}n\mathclose{>}=0$ is satisfied;
such a novel band gap is quite different from a conventional {\em
Bragg reflection gap} because it appears due to completely
different physics of wave reflection. In particular, the periodic
structures with zero averaged refractive index demonstrate a
number of unique properties of the beam transmission observed as
strong beam modification and reshaping~\cite{Shadrivov:2003-3820:APL} being also
insensitive to disorder that is symmetric in the random
variable~\cite{Li:2003-83901:PRL}.

In this paper, we study both {\em transmission properties} and
{\em defect-induced tunability} of the left-handed photonic band gap structures created by alternating slabs of positive and negative refractive
index materials with an embedded defect taking into account realistic parameters, dispersion and losses of the left-handed material. We consider a band-gap structure schematically shown in Fig.~\rpict{geom1}. First, we study the properties of the left-handed material as a composite structure made of arrays of wires and split-ring resonators (see the insert
in Fig.~1) and derive the general results for the effective
dielectric permittivity and magnetic permeability. Second, we study the transmission of electromagnetic waves through the layered structure consisting of alternating slabs of composite left-handed metamaterial using the calculated effective parameters. We assume that the structure includes a defect layer (see Fig.~\rpict{geom1}) that
allows tunability of the transmission  near the defect frequency.
Using the transfer-matrix method, we describe the defect-induced
localized states in such a structure and reveal that the defect
modes may appear in different parameter regions and for both
$\mathopen{<}n\mathclose{>}=0$ and Bragg scattering band gaps.
Depending on the defect parameters, the maximum transmission can
be observed in all or just some spectral band gaps of the structure.
We demonstrate that the frequency of the defect mode  is less
sensitive to manufacturing disorder for the larger defect layer.
Finally, we perform two-dimensional FDTD numerical simulations
based on the microscopic parameters of the left-handed material
and study the temporal evolution of the transmitted and reflected fields.

\pict{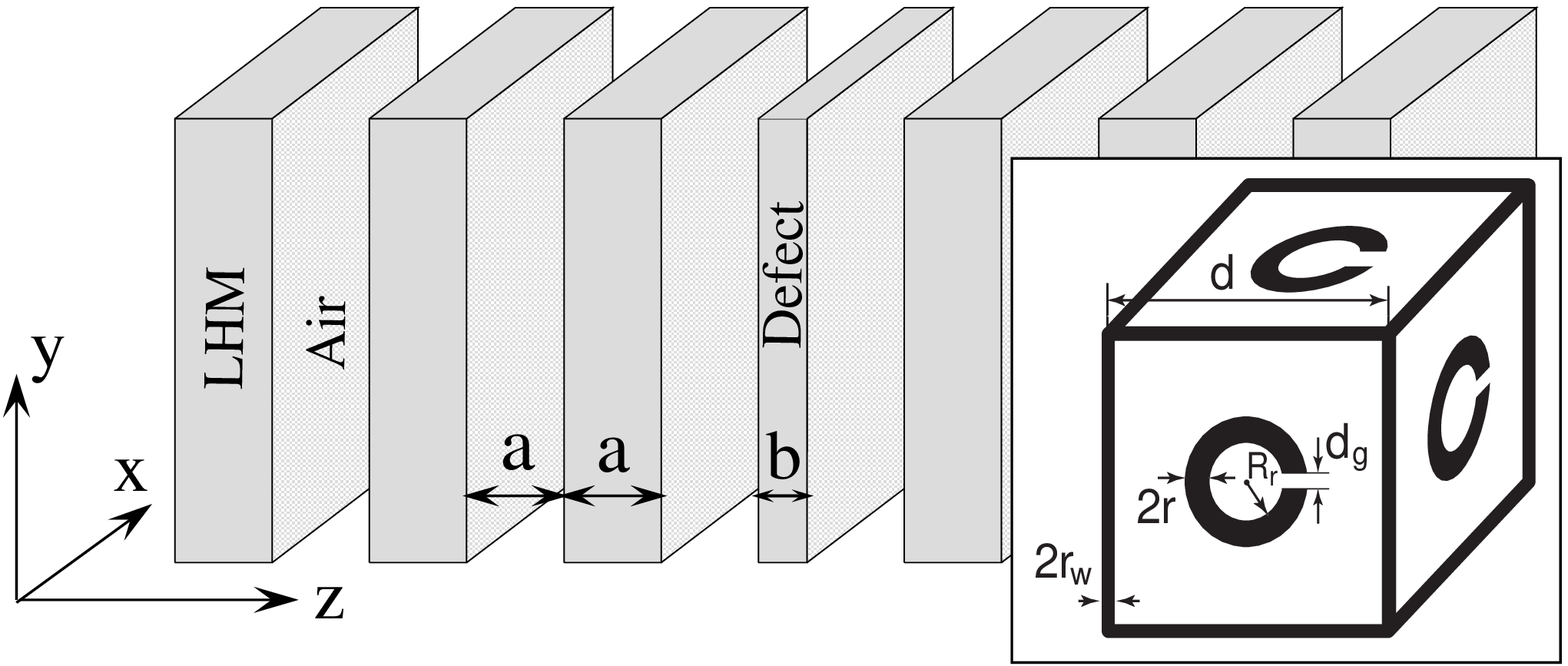}{geom1}{Schematic of a multilayered structure
consisting of alternating metamaterial slabs and air. The
inset shows the unit cell of the metamaterial structure.}

\section{Metamaterial characteristics}

We assume that the left-handed metamaterial is created by a three-dimensional
lattice of wires and split-ring resonators (SRRs), as shown in the
inset of Fig.~\rpict{geom1}. According to the results of the derivation presented in Refs.~\cite{Pendry:1996-4773:PRL,Zharov:2003-37401:PRL}, the main
contribution to the effective dielectric permittivity of this
structure is given by the wires, whereas the magnetic response is
determined by SRRs. Although a three-dimensional lattice of wires contains closed circuits, we neglect their contribution to the magnetic permeability, because this effect is non-resonant, and, therefore, is weak. The effective dielectric permittivity can
be obtained in the form~\cite{Pendry:1996-4773:PRL, Zharov:2003-37401:PRL}
\begin{equation}\leqt{permittivity}
\epsilon \left( \omega \right) = 
	1 - \frac{\omega_p^2}{\omega(\omega-i\gamma_{\epsilon})},
\end{equation}
where $\omega_p \approx (c/d) [2\pi/\ln{(d/r_w)}]^{1/2}$ is the
effective plasma frequency, $\gamma_{\epsilon} = c^2/2\sigma S
\ln{(d/r_w)}$, $\sigma$ is the wire conductivity, $S$ is the effective cross-section of a wire, $S = \pi r_w^2$, for
$ \delta > r_w$, and $S \approx \pi \delta(2r-\delta)$, for $\delta
<r_w$, where $\delta = c/\sqrt{2\pi\sigma\omega}$ is the skin-layer
thickness.

To calculate the effective magnetic permeability of the lattice of SRRs we write its magnetization $M$ in the form (see also Ref.~\cite{Pendry:1999-2075:ITMT})
\begin{equation}\leqt{magnetization}
M = (n_m/2c) \pi R_r^2 I_r,
\end{equation}
where  $n_m = 3/d^3$ is the number of SRRs per unit cell, $R_r$ is
the SRR radius (see the insert in Fig.~\rpict{geom1}), $I_r$ is the current in
the SRR. We assume that SRR is an effective oscillatory
circuit with inductance $L$ and resistance $R$ of the wire, and
capacity $C$ of the SRR slot. In this circuit the electromotive force in this
circuit due to an alternating magnetic field of the propagating wave.
Under these assumptions, the evolution of the current $I_r$ in single SRR is governed by the equation
\begin{equation}\leqt{current}
L\frac{d^2I_r}{dt^2} +
R\frac{dI_r}{dt} +
\frac{1}{C}I_r = \frac{d{\cal E}}{dt} ,
\end{equation}
with \[ {\cal E} = - \frac{\pi R_r^2}{c} \frac{d H^{\prime}}{dt},
\]
where $H^{\prime}$ is the acting (microscopic) magnetic field,
which differs from the average (macroscopic) magnetic field.
We describe the SRR array as a system of magnetic dipoles, which is
valid when the number of SRRs in the volume $\lambda^3$ is big enough, and use  the Lorenz-Lorentz relation between the microscopic and macroscopic magnetic fields~\cite{Landau:1963:Electrodynamics}
\begin{equation}\leqt{L-L}
H^{\prime} = H + \frac{4\pi}{3} M = B - \frac{8\pi}{3} M.
\end{equation}
As a result, from Eqs.~(\ref{eq:magnetization})--(\ref{eq:L-L}) we obtain {\em the effective magnetic permeability} of the structure in the form,
\begin{equation}\leqt{permeability}
\mu \left( \omega \right) = 
	1 + \frac{F \omega^2}{\omega_0^2 - \omega^2 \left(1+F/3\right) +
	i\omega \gamma},
\end{equation}
where $F = 2\pi n_m(\pi R_r^2/c)^2/L$, $\omega_0^2 = 1/LC$, and $\gamma = R/L$. Inductance $L$, resistance $R$, and capacitance $C$ are given by the following results (see, e.g., Ref.~\cite{Schwinger:1998:ClassicalElectrodynamics}),
\[
L = \frac{4\pi R_r}{c^2} \left[\ln \left(\frac{8R_r}{r}\right) -
\frac{7}{4} \right], \;\; R = \frac{2\pi R_r}{\sigma S_r}, \;\; C =
\frac{\pi r^2}{4\pi d_g}, 
\]
where $r$ is the radius of the wire that makes a ring, $\sigma$ is conductivity of the wire, $S_r$ is the effective area of the cross-section of SRR wire defined similar to that of a straight wire, and $d_g$ is the size of the SRR slot. We note, that the result for $C$ should hold provided $d_g \ll r$.

Taking the parameters of a metallic composite as $d =
1$ cm, $r_w = 0.05$ cm, $R_r = 0.2$ cm, $r = 0.02$ cm, $d_g = 10^{-3}$ cm,
and its conductivity as $\sigma = 2\cdot10^{19}$ s$^{-1}$, we calculate
the effective frequency dependencies of $\epsilon(\omega)$ and $\mu(\omega)$ according to Eqs.~\reqt{permittivity} and \reqt{permeability}, respectively, and show these dependencies in Fig.~\rpict{eps_mu}. The resonance frequency appears near $5.82$ GHz, and the region of simultaneously negative $\epsilon$
and $\mu$ is between $5.82$ GHz and $5.96$ GHz. The imaginary part
of the magnetic permeability, which determines effective losses in a left-handed
material, is larger near the resonance.

\pict{fig02.eps}{eps_mu}{Frequency dependence of the real part of the
dielectric permittivity Re($\epsilon$) (solid), and the real part of
magnetic permeability Re($\mu$)(dashed).}

\section{Transmission and defect modes}

We consider now a band gap structure created by seven alternating
left-handed and dielectric slab pairs, as shown in Fig.~\rpict{geom1}. The number of slabs is chosen to keep losses in the structure at a reasonably low
level, still having visible the effects of periodicity. We assume
that the periodic structure is created by slabs of the width
$a$, made of the left-handed composite described above, and separated by
air ($\epsilon_a = 1$, $\mu_a = 1$). The middle layer of the
left-handed material is assumed to have a different thickness, $b = a
(1+\Delta)$, in order to describe {\em a structural defect.}

To study the transmission characteristics of such a band-gap structure,
we consider a normal-incidence plane wave for the scalar electric
field described by the Helmholtz-type equation,
\begin{equation} \leqt{Helm}
   \left[ \frac{d^2}{dz^2}
   + \frac{\omega^2}{c^2} \epsilon(z)\mu(z) 
   - \frac{1}{\mu(z)} \frac{d \mu}{d z} \frac{d}{dz}\right] E
   = 0,
\end{equation}
where $\epsilon(z)$ and $\mu(z)$ are the dielectric permittivity
and magnetic permeability in the whole space. 

Before analyzing the transmission properties of a finite layered structure, first we study the corresponding {\em infinite} structure without defects and
calculate its band gap spectrum. In an infinite periodic structure, propagating waves in the form of {\em the Floquet-Bloch modes} satisfy the periodicity condition, $E(z + 2a) = E(z) \exp( 2i \,a\, K_b)$, where $K_b$ is the Bloch wavenumber. The values of $K_b$ are found as solutions of the standard eigenvalue equation for a two-layered periodic structure (see, e.g., Ref.~\cite{Wu:2003-235103:PRB}),
\begin{eqnarray} \leqt{TransferM}
   2 \cos(K_b 2a) = 2 \cos{\left[(k_{r} + k_{l}) a \right]} -\nonumber\\
   - \left( \frac{p_r}{p_l} + \ \frac{p_l}{p_r} - 2 \right)
            \cdot \sin{\left( k_{r} a  \right)}
            \cdot \sin{\left( k_{l} a  \right)},
\end{eqnarray}
where $p_r = 1$, $p_l = \sqrt{\epsilon/\mu}$, $k_{r} =
\omega/c$ and $k_{l} = \omega/c\sqrt{\epsilon \mu}$ are the
wavevectors in air and left-handed slabs, respectively. For real
values of $K_b$, the Bloch waves are propagating; complex values of $K_b$
indicate the presence of {\em band gaps}, where the wave
propagation is suppressed. The spectral gaps appear when the
argument of the cosine function in Eq.~\reqt{TransferM} becomes
the whole multiple of $\pi$, and no real solutions for $K_b$ exist. These gaps are usually termed as Bragg gaps. The presence of the left-handed material in the structure makes it possible for the argument to vanish, so that the wave propagation becomes prohibited in this case as well, thus creating an {\em additional band gap}, which do not exist in conventional periodic structures. As a matter of fact, the condition $|k_{r}| = |k_{l}|$ corresponds to {\em the zero average refractive index}, $\mathopen{<}n\mathclose{>}=0$, as discussed in Refs.~\cite{Nefedov:2002-36611:PRE, Li:2003-83901:PRL, Ruppin:2003-494:MOTL, Shadrivov:2003-3820:APL, Wu:2003-235103:PRB}. However, the inherent feature of the left-handed materials is their frequency dispersion, so
that the condition $|k_{r}| = |k_{l}|$ defines a characteristic
frequency $\omega^*$ at which the indices of refraction in both
the media compensate each other. In a sharp contrast to
the conventional Bragg reflection gaps, the position of this additional
$\mathopen{<}n\mathclose{>}=0$ gap  in the frequency domain does
not depend on the optical period of the structure.

\pict{fig03.eps}{thickness}{Transmission coefficient of a finite periodic
structure with seven layers of left-handed material vs. the normalized slab
thickness $a/\lambda$, where $\lambda = 2\pi c/\omega^*$ and the frequency 
$\omega = \omega^*$ corresponds to the condition
$\mathopen{<}n\mathclose{>}=0$. }

For the parameters of the left-handed media described above,
the frequency $\omega^*$ at which the average refractive index
vanishes is found as $\omega^* \approx 2\pi \cdot 5.8636\times10^9$ $s^{-1}$.
Importantly, the transmission coefficient calculated at $\omega =
\omega^*$ for the seven-layer structure shows a characteristic
resonant dependence as a function of the normalized slab
thickness $a/\lambda$, where $\lambda = 2\pi c/\omega^*$, as shown
in Fig.~\rpict{thickness}. The transmission maxima appear in the
$\mathopen{<}n\mathclose{>}=0$ gap, when the slab thickness $a$ coincides
with a whole multiple of a half of the wavelength. The width of
the transmission peaks decreases with the increase of the number
of layers in the structure. The transmission maxima decrease
with increasing thickness of the structure due to losses in the
left-handed material which become larger for thicker slabs. One of
the interesting features of the $\mathopen{<}n\mathclose{>}=0$ gap
is that the transmission coefficient can vanish even for very
small values of the slab thickness. This property can be employed
to create {\em effective mirrors} in the microwave region operating
in this novel $\mathopen{<}n\mathclose{>}=0$ gap which can be {\em effectively
thinner than the wavelength} of electromagnetic waves.

\pict{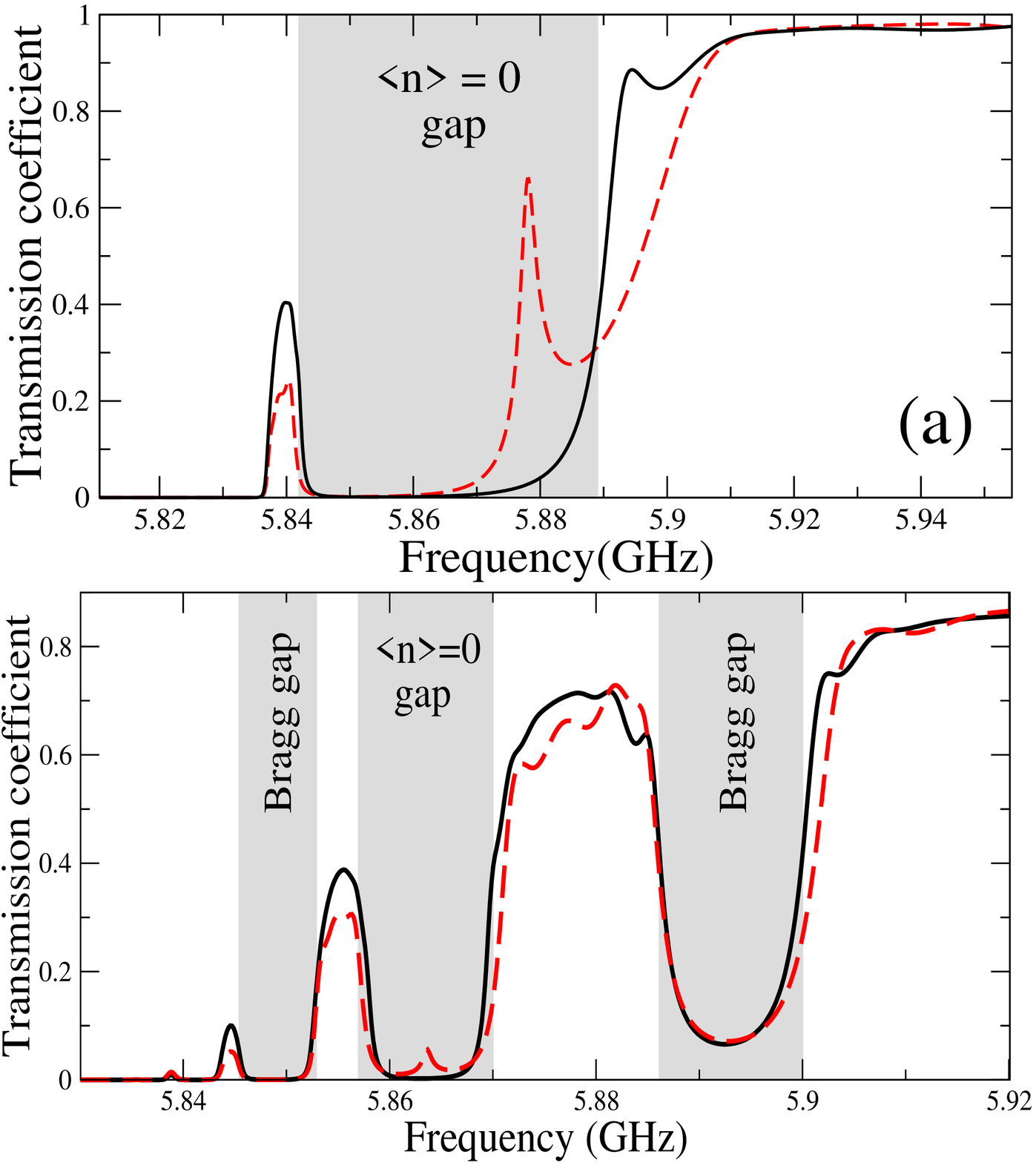}{band-gaps}{Transmission coefficient of the
left-handed band-gap structure vs. the wave frequency. (a) The
structure with the period $a = 0.25\lambda$, without (solid) and
with (dashed) defect layer ($\Delta = -0.8$). (b) The structure with the
period $a = 1.25\lambda$ without (solid) and with (dashed) defect layer
($\Delta = -0.6$).}

\pict{fig05.eps}{def_position}{Frequency spectrum of the defect
modes  as a function of the normalized defect size $\Delta$ in the
left-handed band-gap structure with the period $a = 1.25\lambda$.}

The transmission coefficient is shown in Fig.~\rpict{band-gaps}(a,b) as a function of the frequency, for two structures with different values of the period $a$. For the quarter-wavelength slabs [see Fig.~\rpict{band-gaps}(a)], the only visible band gap is the $\mathopen{<}n\mathclose{>}=0$ gap
centered near $\omega^*$ where both $\epsilon$ and $\mu$ are
negative. When the structure has a defect, the transmission peak
associated with the excitation of the defect mode appears inside
the $\mathopen{<}n\mathclose{>}=0$ gap as shown by the dashed curve. For the structure with thicker slabs, e.g. for $a = 1.25 \lambda$ [see Fig.~\rpict{band-gaps}(b)] the $\mathopen{<}n\mathclose{>}=0$ gap becomes narrower but it remains centered near the frequency $\omega^*$. The transmission
coefficient of this second band-gap structure shows, in addition to the
$\mathopen{<}n\mathclose{>}=0$ gap, two Bragg scattering gaps. Due
to the increased losses in this second band-gap structure, where slabs are thicker than in the structure corresponding to Fig.~\rpict{band-gaps}(a),  the
effects of the resonant transmission at the defect mode become
less visible. Moreover, for the parameters we consider here the defect
mode appears only in the $\mathopen{<}n\mathclose{>}=0$ gap,
whereas {\em it does not appear} in the Bragg gaps. For larger slab
thickness, higher-order Bragg gaps may appear in the frequency
range where the composite material is left-handed.

\pict{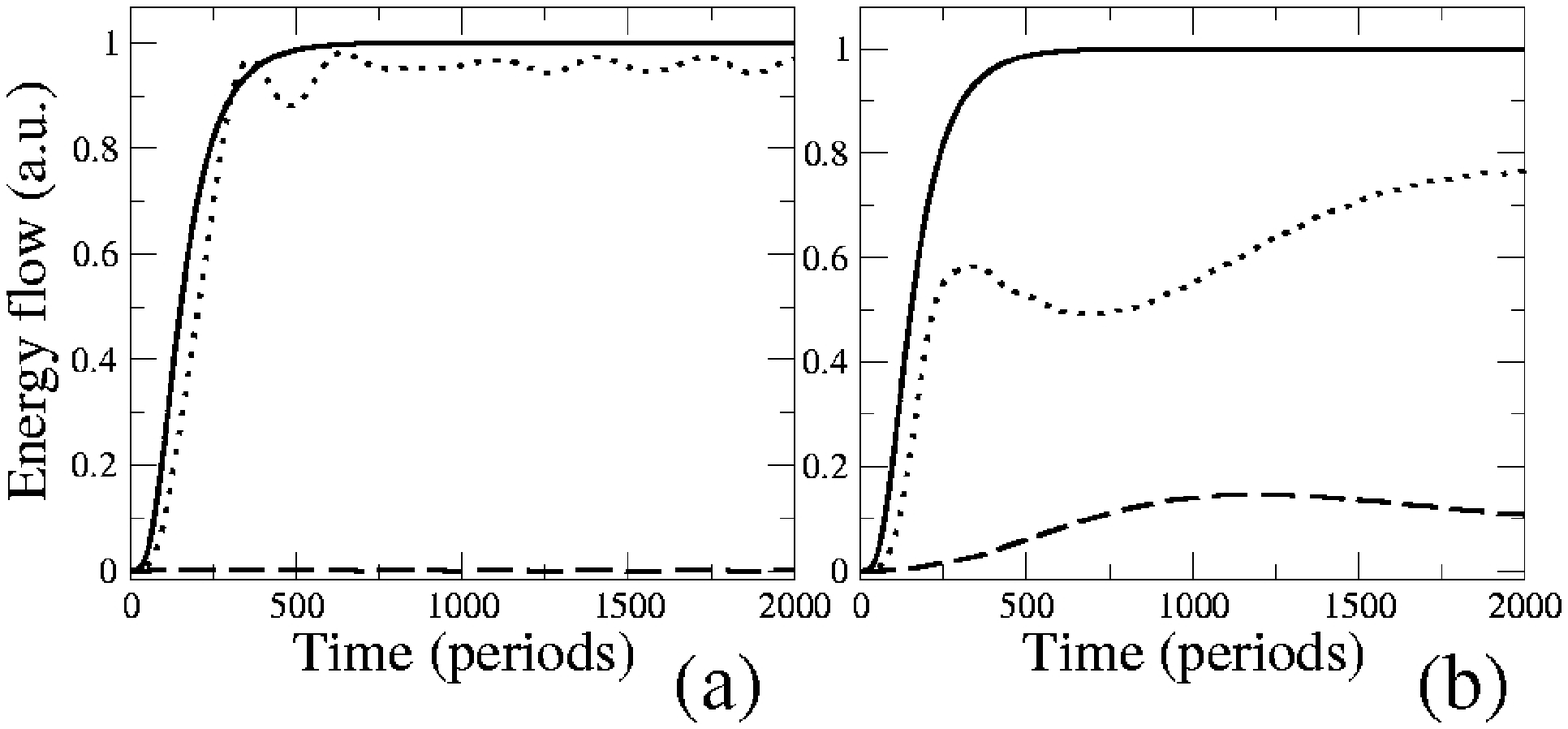}{power_tr}{Numerical FDTD simulation results showing relaxation processes in a band-gap structure with a defect. Solid -- incident energy flow, dashed -- transmitted energy flow, dotted -- reflected energy flow. Parameters are: $a = 0.25\lambda$, $\Delta = -0.8$. (a) Defect mode is not excited, $\omega = 2 \pi 5.86 \times 10^9$ $s^{-1}$; (b) Defect mode is
excited, $\omega = 2\pi 5.878 \times 10^9$ $s^{-1}$.}
\pict{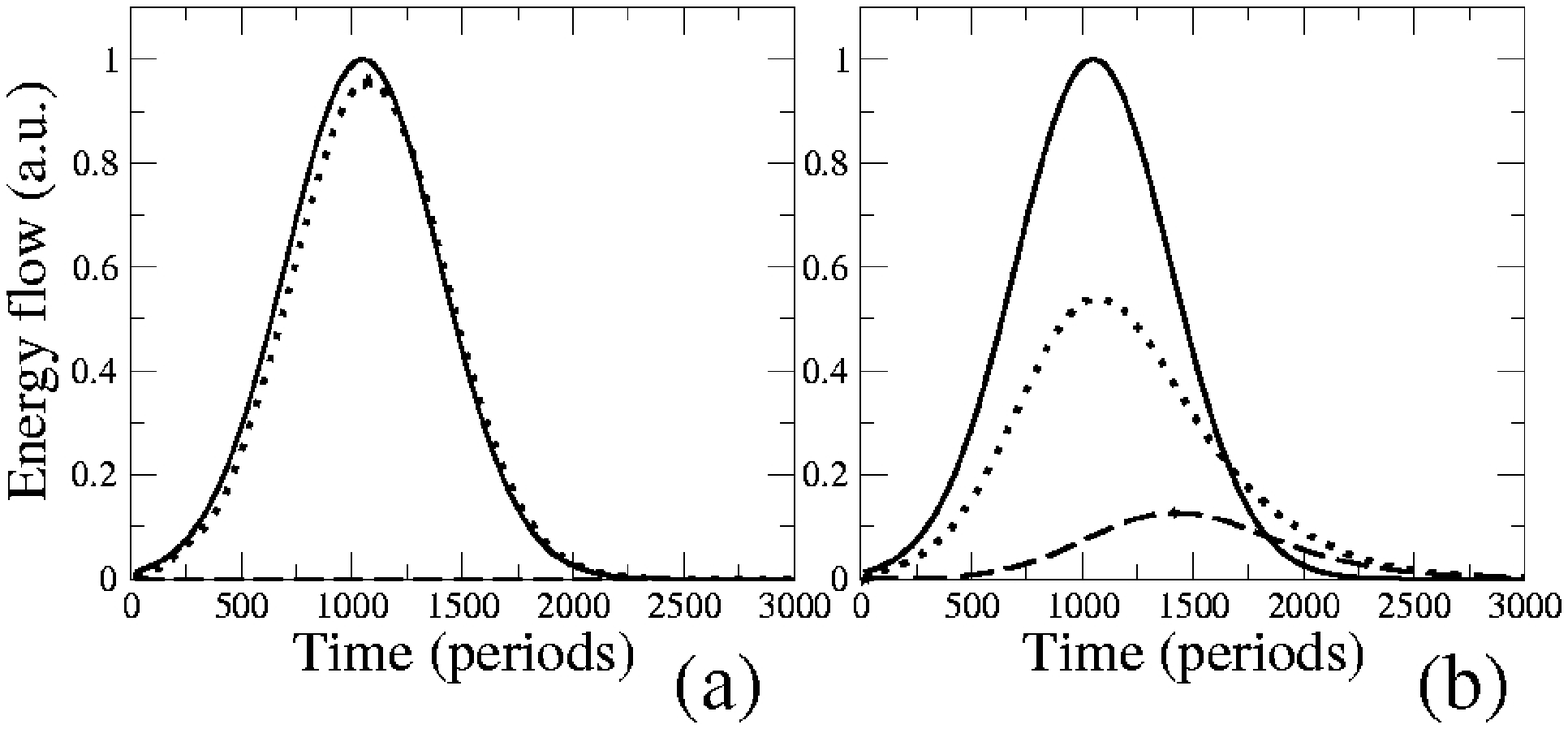}{power_tr_pulse}{Numerical FDTD simulation results for the pulse scattering by a periodic structure with defect. Solid -- incident energy flow, dashed -- transmitted energy flow, dotted -- reflected energy flow.
Parameters are: $a = 0.25\lambda$, $\Delta = -0.8$. (a) Defect
mode is not excited, $\omega = 2 \pi 5.86 \times 10^9$ $s^{-1}$; (b) Defect mode is
excited, $\omega = 2\pi 5.878 \times 10^9$ $s^{-1}$.}
In Fig.~\rpict{def_position}, we show the frequency spectrum of
the defect modes for the structure with $a = 1.25\lambda$ as a
function of the normalized defect size $\Delta$. We
notice a number of important features: (i)  the defect modes do
not always appear simultaneously in all gaps, and (ii) the
slope of the curves in Fig.~\rpict{def_position} decreases with
the growth of the defect thickness. As a result, the eigenfrequencies of the modes introduced by thicker defect layers can be more stable to disorder introduced by manufacturing. These novel features seem to be important for tunable properties of the layered structures because the different modes allow to access different types of band gaps. 

\pict{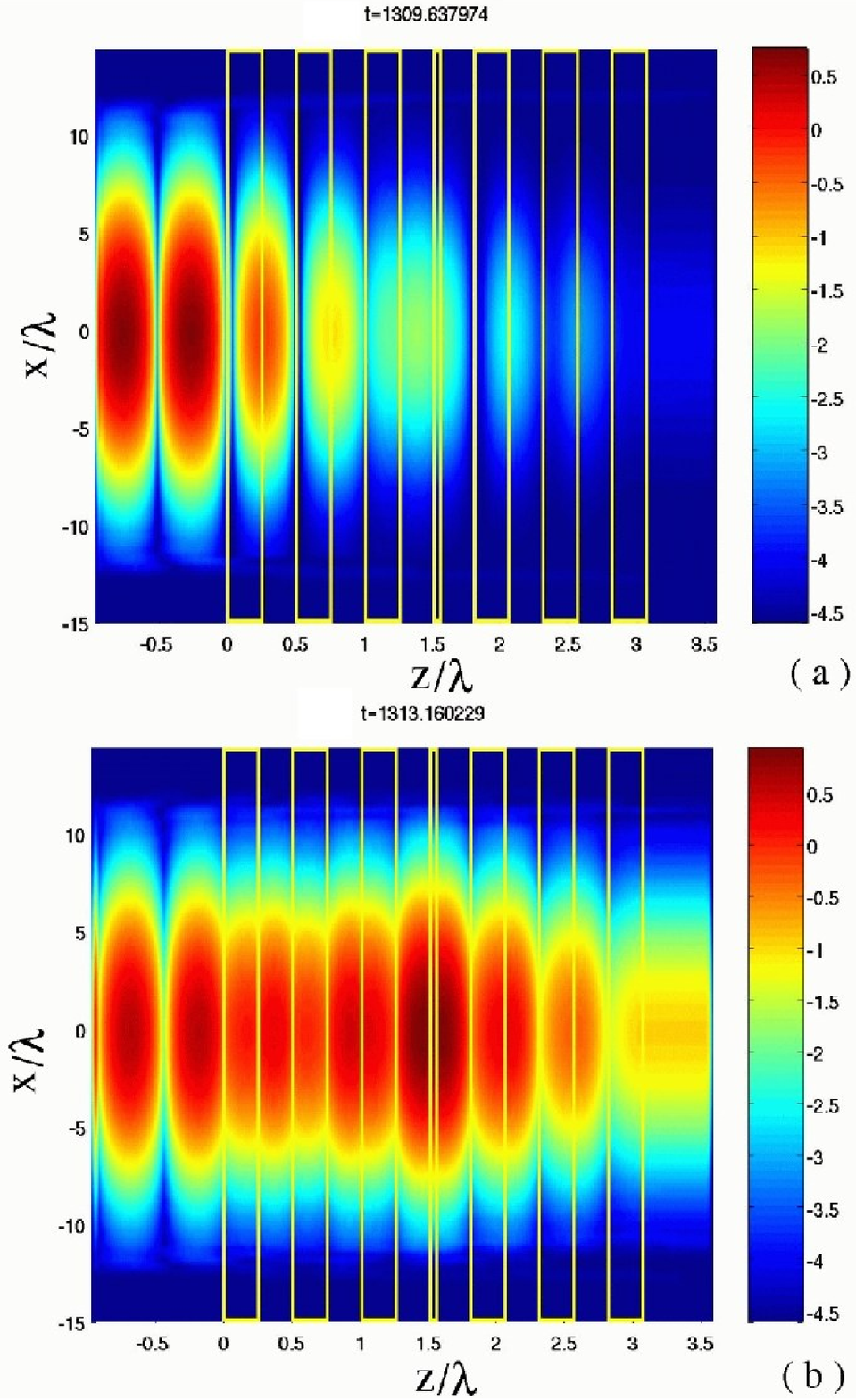}{fields}{Results of the numerical FDTD simulations for the amplitude of the magnetic field in a two-dimensional structure (natural logarithm scale). Boxes show positions of the left-handed slabs, $a =
0.25\lambda$, $\Delta = -0.8$. (a) Defect mode is not excited,
$\omega = 2 \pi 5.86 \times 10^9$ $s^{-1}$; (b) Defect mode is excited, $\omega =
2\pi 5.878 \times 10^9$ $s^{-1}$. }

\section{Numerical FDTD simulations}

In order to analyze the temporal evolution of the transmitted
fields and the beam scattering under realistic conditions, we
perform two-dimensional FDTD numerical simulations of the beam
propagation through the left-handed periodic structure of seven
layers with a defect. We consider TM-polarized Gaussian beam of the width $20\lambda$ propagating towards the structure with the period $a = 0.25\lambda$; such a
structure corresponds to the transmission coefficient shown in
Fig.~\rpict{band-gaps}(a) by a dashed line. First, we choose the
frequency of the incident field in the
$\mathopen{<}n\mathclose{>}=0$ gap. The temporal evolution of the
energy flows (integrated over the transverse dimension) for the
incident, transmitted, and reflected waves is shown in
Fig.~\rpict{power_tr}(a), clearly indicating that the transmission
through such a structure is negligible. When the frequency of the
incident field coincides with that of the defect mode, a
significant amount of the energy is transmitted through the
structure [see Fig.~\rpict{power_tr}(b)]. The
relaxation time of the beam transmission through the structure is estimated as $10^3$ periods (approximately $170$ ns).

Results of FDTD simulations for the {\em pulse scattering} from the structure with $a = 0.25\lambda$ are shown in Figs.~\rpict{power_tr_pulse} (a,b) as the temporal dependence of the incident, reflected and transmitted energy flows. One can clearly see significant transmission, when the carrier frequency of the pulse coincides with the eigen frequency of the defect mode [see Figs.~\rpict{power_tr_pulse} (a,b)]. 

An example of the field distribution in the structure with the slab size $a = 0.25\lambda $ is shown in Figs.~\rpict{fields} (a,b) for two
different regimes. In Fig.~\rpict{fields}(a), the frequency
corresponds to low transmission in Fig.~\rpict{band-gaps}(a), when
no defect mode is excited. Figure~\rpict{fields}(b) demonstrates the
field distribution in the structure with an excited defect mode
and enhanced transmission.

\section{Conclusions}

We have studied the transmission properties of periodic structures made of a left-handed metamaterial and air. Using realistic parameters of the metamaterial derived from the microscopic approach, we have calculated the
band-gap spectrum of an infinite one-dimensional structure with
and without defects, demonstrating the existence of band gaps of
two different types, the conventional Bragg scattering gaps and a
novel $\mathopen{<}n\mathclose{>}=0$ gap. We have analyzed the
properties of the defect modes in a finite periodic structure with
a structural defect and demonstrated that, depending on the defect
size, the defect modes appear in all or just some of the band gaps
allowing to access different gaps selectively. In addition, we
have performed two-dimensional numerical FDTD simulations of the propagation of electromagnetic waves in such structures and have studied the temporal dynamics of the beam transmission and reflection. We have demonstrated that the excitation of defect modes can enhance substantially the wave transmission through the structure.

The authors thank Michael Feise for collaboration, and acknowledge a partial support from the Australian Research Council.

\end{sloppy}
\end{document}